\documentclass[12pt]{article} 

\newcommand\beq{\begin{equation}}
\newcommand\eeq{\end{equation}}

\date{}
\usepackage[final]{graphicx}
\usepackage{amsmath}
\usepackage{amsfonts}   
\usepackage{amssymb}
\begin{document}


\title{Radiation Properties of One-Dimensional Quasirandom Antenna Arrays Based on Rudin-Shapiro Sequences}

\author{Vincenzo Galdi,  
Vincenzo Pierro, Giuseppe Castaldi,\\ Innocenzo M. Pinto,  and
Leopold B. Felsen 
\thanks{
V. Galdi, V. Pierro, G. Castaldi, and I.M. Pinto are with the Waves Group, Department of
Engineering, University of Sannio, I-82100 Benevento, Italy (e-mail: vgaldi@unisannio.it, pierro@unisannio.it, castaldi@unisannio.it, pinto@sa.infn.it).\protect\\
L.B. Felsen is with the Department of Aerospace and Mechanical Engineering and the
Department of Electrical and Computer Engineering, Boston University (part-time), Boston, MA 02215 USA. He is also University Professor Emeritus, Polytechnic University, Brooklyn,
NY 11201 USA (e-mail: lfelsen@bu.edu).}}

\maketitle

\begin{abstract}
The development of exotic new materials, such as metamaterials, has created strong interest within the electromagnetics (EM) community for possible new phenomenologies and device applications, with particular attention to periodicity-induced phenomena, such as photonic bandgaps (PBG). Within this context, motivated by the fairly recent discovery in X-ray crystallography of ``quasicrystals,'' whose diffraction patterns display unusual characteristics that are associated with ``aperiodic order,'' we have undertaken a systematic study of how these exotic effects manifest themselves in the radiation properties of aperiodically-configured antenna arrays. The background for these studies, with promising example configurations, has been reported in a previous publication [V. Pierro {\em et al.}, {\em IEEE Trans. Antennas Propagat.}, vol. 53, No. 2, pp. 635--644, Feb. 2005]. In the present paper, we pay attention to various configurations generated by Rudin-Shapiro (RS) sequences, which constitute one of the simplest conceivable examples of {\em deterministic aperiodic} geometries featuring {\em quasirandom} (dis)order. After presentation and review of relevant background material,
the radiation properties of one-dimensional RS-based antenna arrays are analyzed, followed by illustrative numerical parametric studies to validate the theoretical models. Design parameters and potential practical applications are also given attention.
\end{abstract}

\markboth{GALDI et al.: Radiation Properties of Quasirandom Antenna Arrays...}{} 

\section{Introduction}
\label{intro}
In an ongoing plan of investigations \cite{Tiling_AP}, \cite{Fibonacci_AP}, we have recently been concerned with the study of the radiation properties of structures with {\em aperiodic order}. Interest in this type of geometries is physically motivated by the recent discovery of ``quasicrystals'' \cite{Shechtman}, \cite{Levine}, i.e., certain metallic alloys displaying X-ray diffraction properties that are known to be {\em incompatible} with spatial periodicity. In electromagnetic (EM) engineering, this has so far produced interesting applications in the field of photonic bandgap ``quasicrystal'' devices (see the brief summary and references in \cite{Tiling_AP}).

We have initiated our series of prototype studies with the aim of exploring the radiation properties of selected  categories of  aperiodic geometries, representative of the various hierachial types and degrees of aperiodic order in the ``gray zone'' that separates {\em perfect periodicity} from {\em absolute randomness}. To this end, we started in \cite{Tiling_AP} with exploring the basic radiation properties of two-dimensional (2-D) antenna arrays based on certain representative categories of ``aperiodic tilings'' (see \cite{Grunbaum}, \cite{Senechal1} for an introduction to the subject). In parallel, we also began investigating the radiation properties of selected categories of 1-D {\em aperiodic sequences}, which are more easily amenable to explicit {\em analytic parameterization} and can thus provide better insight into the relevant wave-dynamical phenomenologies. Within this framework, we explored in \cite{Fibonacci_AP} the radiation properties of 1-D {\em quasiperiodic} antenna arrays based on the {\em Fibonacci sequence}, which admit analytic parameterization in terms of {\em generalized Poisson summation} formulas. In this paper, we procede along the same route, but focus on the {\em opposite} end of the above-mentioned aperiodic-order ``gray zone'', namely, the {\em quasirandom} type of (dis)order. One of the simplest and most intriguing examples is provided by the so-called {\em Rudin-Shapiro (RS) sequences} \cite{Shapiro1}--\cite{Brillhart1}. These sequences were developed in a pure-mathematics
context by H. S. Shapiro and W. Rudin during the 1950's, pertaining to some extremal problems in harmonic analysis. More recently, RS sequences have found important practical applications as signal processing tools for {\em spread spectrum} communication and encryption systems \cite{Dixon}, \cite{Cour}. Concerning their wave-dynamical properties, available studies are principally focused on the transmission characteristics of 1-D RS-based dielectric multilayers (see, e.g., \cite{Vasco}). 

In line with our stated interest in the radiation properties of aperiodically-parameterized antenna arrays, we concentrate in the present paper on 1-D RS-based antenna array configurations. Our implementation via analytic and numerical studies makes use of well-established spectral and statistical properties of RS sequences.

Accordingly, the remainder of the paper is laid out as follows. Section \ref{formulation} contains the problem formulation, with description of the geometry and review of the basic properties of RS sequences. Section \ref{Array} deals with the radiation properties of representative classes of RS-based antenna arrays; in this connection, potential applications to {\em omnidrectional} and {\em thinned} arrays are proposed and discussed. Conclusions are provided in Section \ref{Conclusions}.

\section{Background and Problem Formulation}
\label{formulation}

\subsection{Geometry}
\label{Geom}
\begin{figure} 
\begin{center}
\includegraphics[width=10cm]{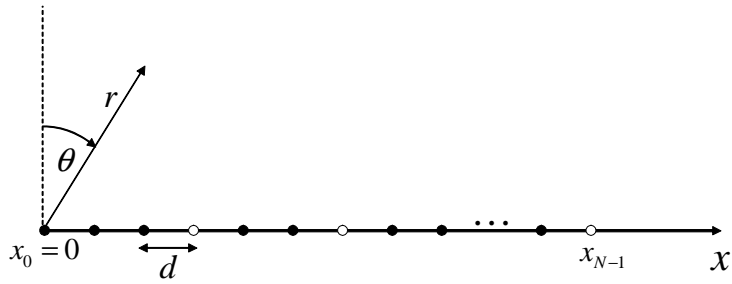} 
\end{center}
\caption{Problem geometry: A 1-D array of $N$ identical antenna elements with uniform inter-element spacing $d$ is considered, with antenna element positions at $x_n=n d, n=0,...,N-1$, and uniform inter-element phasing $\eta$. Black and white dots denote the two possible values of the real excitation amplitudes $I_n$ in (\ref{eq:In}), related to the RS sequence (see Sec. \ref{Rud_Shap}). 
Also shown is the array-end-centered polar $(r,\theta)$ coordinate system.} 
\label{Figure1} 
\end{figure}

The problem geometry, a 1-D array of $N$ identical antenna elements with uniform inter-element spacing $d$, is sketched in Fig. \ref{Figure1}. Without loss of generality, the array is assumed to be placed on the $x$-axis, with radiators at $x_n=n d, n=0,...,N-1$; in view of the rotational symmetry around the $x$-axis, we adopt the 2-D polar $(r,\theta)$ coordinate system. Neglecting inter-element coupling, and for an implied $\exp(j \omega t)$ time dependence, the $n$-th antenna element complex excitation is denoted by
\beq
{\tilde I}_n=I_n\exp(-jk_0nd\eta),
\label{eq:In}
\eeq
where $I_n$ is a real excitation amplitude which, by assumption, can take on only two values (denoted by black and white dots in Fig. \ref{Figure1}) related to the RS sequence described below (Sec. \ref{Rud_Shap}). Moreover,
$k_0=\omega\sqrt{\epsilon_0\mu_0}=2\pi/\lambda_0$ is the free-space wavenumber (with $\lambda_0$ being the wavelength), and $\eta$ describes the inter-element phasing. For observation distances $r\gg  2(Nd)^2/\lambda_0$ (Fraunhofer region), the array is characterized by the ``array factor'' \cite{Mailloux}
\begin{eqnarray}
F_s(\xi)&=&\sum_{n=0}^{N-1}I_n\exp\left(
j2\pi n\xi
\right)\nonumber\\
&=&\sum_{n=0}^{N-1}I_n \exp\left[j k_0 n d \left(\sin\theta-\eta\right)
\right]=F_p(\theta),
\label{eq:AF}
\end{eqnarray}
where $\xi=d(\sin\theta-\eta)/\lambda_0$ is a nondimensional angular spectral variable. In what follows, we shall refer to $|F_s(\xi)|^2$ as the ``radiation spectrum'', and to $|F_p(\theta)|^2$ as the ``radiation pattern''. The 1-D results discussed here can be extended to 2-D geometry-separable configurations via conventional pattern multiplication \cite{Mailloux}.

\subsection{Rudin-Shapiro Sequences}
\label{Rud_Shap}
Rudin-Shapiro (RS) sequences are two-symbol aperiodic sequences with {\em quasirandom} character \cite{Shapiro1}--\cite{Brillhart1}. In what follows, 
their basic properties are briefly reviewed, with emphasis on those aspects that are directly relevant to the antenna array application in Sec. \ref{Array} (see \cite{Queffelec}--\cite{Fogg} for more details).
In its basic form, a RS sequence is generated from the {\em two-symbol alphabet} ${\cal A}_a=\left\{-1,1\right\}$ via the simple recursive rule
\beq
\alpha_0=1,~~
\alpha_{2n}=\alpha_n,~~
\alpha_{2n+1}=(-1)^n\alpha_n.
\label{eq:ARS}
\eeq
Thus, for instance, the first ten symbols in the sequence are
\beq
1~~1~~1~ -\!1~~1~~1~ -\!1~~1~~1~~1.
\eeq
It can be shown that, in the limit of an {\em infinite} sequence, the two symbols have the same statistical frequency of occurrence \cite{Berthe}. Throughout the paper, the sequence $\left\{\alpha_n\right\}$ in (\ref{eq:ARS}) will be referred to as an {\em alternate} RS (ARS) sequence. This sequence can also be generated via a two-step procedure involving a substitution rule defined on a four-symbol alphabet ${\cal A}_s=\left\{A,B,C,D\right\}$, 
\begin{subequations}
\beq
A\rightarrow AB,~~B\rightarrow AC,~~C\rightarrow DB,~~D\rightarrow DC,
\eeq
and a final projection $\Pi :{\cal A}_s\rightarrow{\cal A}_a$,
\beq
\Pi(A)=1,~~\Pi(B)=1,~~\Pi(C)=-1,~~\Pi(D)=-1.
\eeq
\label{eq:ARS1}
\end{subequations} 
Another possible construction algorithm is based on the use of the so-called {\em RS polynomials} \cite{Cour}, \cite{Axel}, \cite{Brillhart} defined by the recursive rules 
\begin{subequations}
\begin{eqnarray}
P_{m+1}(\xi)=P_m(\xi)+\exp\left(j2\pi2^m\xi\right)Q_m(\xi),~~P_0=1,\\
Q_{m+1}(\xi)=P_m(\xi)-\exp\left(j2\pi2^m\xi\right)Q_m(\xi),~~Q_0=1.
\end{eqnarray}
\label{eq:PQ}
\end{subequations}
It can be verified that $P_m$ and $Q_m$ in (\ref{eq:PQ}) are $(2^m-1)$-th degree trigonometric polynomials in the variable $\exp(j2\pi\xi)$, 
\begin{subequations}
\begin{eqnarray}
P_m(\xi)=\sum_{n=0}^{2^m-1}p_n \exp\left(j2\pi n \xi\right),\\
Q_m(\xi)=\sum_{n=0}^{2^m-1}q_n \exp\left(j2\pi n \xi\right).
\end{eqnarray}
\label{eq:PQ1}
\end{subequations}
The link between RS polynomials and sequences stems from the polynomial coefficients $p_n$ and $q_n$ in (\ref{eq:PQ1}). For $P_m$-type polynomials, the coefficient sequence coincides with the ARS sequence, i.e., $p_n=\alpha_n$, $n=0,...,2^m-1$. For $Q_m$-type polynomials, the coefficient sequence (technically defined as ARS Golay-complementary \cite{Golay}) is obtained in practice by reversing the sign of the second half ($n\ge 2^{m-1}$) of the ARS sequence,
\beq
q_n=
\left\{
\begin{array}{ll}
\alpha_n,~~0\le n\le 2^{m-1}-1,\\
-\alpha_n,~~2^{m-1}\le n\le2^m-1.
\end{array}
\right.
\eeq
The reader is referred to \cite{Axel}, \cite{Allouche}, \cite{Benke} for more details and possible generalizations of ARS sequences. Antenna arrays based on ARS sequences are treated in Sec. \ref{Flat}. 

Another RS sequence of interest for this investigation (see Sec. \ref{Thinned}) is related to the ARS sequence in (\ref{eq:ARS}) via the simple linear transformation,
\beq
\beta_n=\frac{1-\alpha_n}{2}.
\label{eq:BRS}
\eeq
It is readily verified that the sequence $\left\{\beta_n\right\}$ in (\ref{eq:BRS}) is composed of symbols from the alphabet ${\cal A}_b=\left\{0,1\right\}$, and will accordingly be defined as {\em binary} RS (BRS) sequence.

Despite the {\em fully deterministic} and {\em deceptively simple} character of (\ref{eq:ARS}) and (\ref{eq:BRS}), RS sequences exhibit some {\em statistical and spectral} properties that
are remarkably different from those of typical deterministic sequences, and similar in some respects to those of {\em random} sequences. In particular, the correlation properties of the ARS sequence in (\ref{eq:ARS}) resemble those of white-noise-like random sequences \cite{Fogg},
\beq
\lim_{N\rightarrow\infty}\frac{1}{N}
\sum_{n=0}^{N-1}
\alpha_n\alpha_{n+\nu}=\delta_{\nu0},
\label{eq:corr}
\eeq
where $\delta_{pq}$ is the Kronecker delta symbol, $\delta_{pq}=1$ for $p=q$, $\delta_{pq}=0$ otherwise.

From the spectral viewpoint, ARS sequences exhibit interesting {\em extremal} properties. It can be proved that their truncated Fourier series amplitude is bounded by \cite{Rudin1}
\beq
\sqrt{N}\le \sup_{0\le \xi<1}
\left|
\sum_{n=0}^{N-1}
\alpha_n\exp\left(j 2\pi n \xi\right)\right|\le (2+\sqrt{2})\sqrt{N}.
\label{eq:Fbound}
\eeq
Note that while the lower bound in (\ref{eq:Fbound}) holds for {\em any} sequence $\left\{\alpha_n\right\}$ generated from the alphabet ${\cal A}_a=\left\{-1,1\right\}$ \cite{Fogg}, the upper bound is peculiar of ARS sequences. For 
a general sequence $\{\alpha_n\}$ generated from the alphabet ${\cal A}_a$, the upper bound in (\ref{eq:Fbound}) would be $N$ (with the equality sign holding, e.g., for {\em periodic} sequences) \cite{Queffelec}, \cite{Fogg}. On the other hand, for {\em random} sequences in ${\cal A}_a$, the upper bound typically behaves like $\sqrt{N\log N}$ \cite{Queffelec}. 

\begin{figure} 
\begin{center}
\includegraphics[width=12cm]{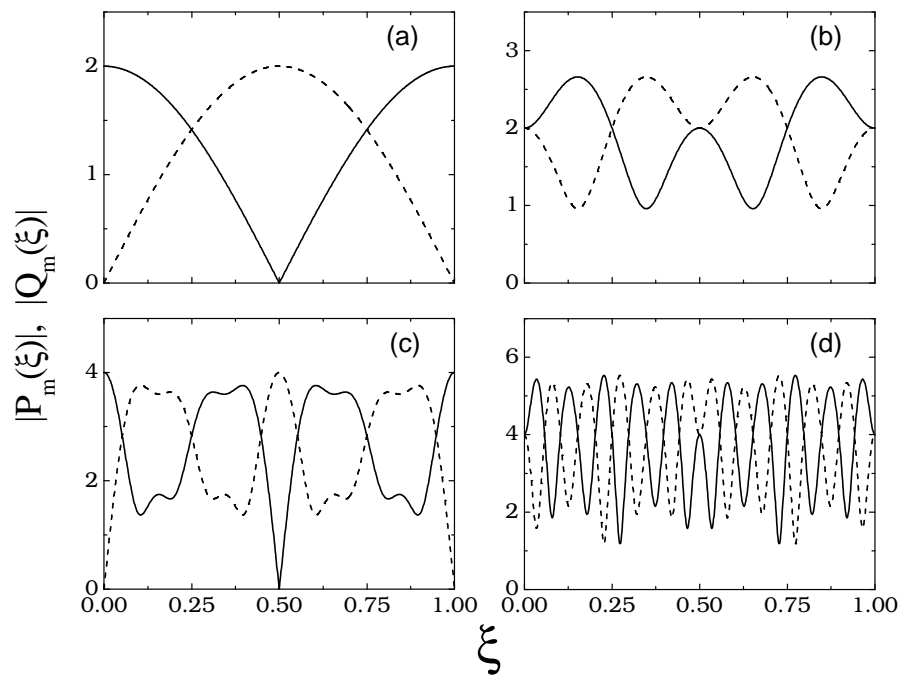} 
\end{center}
\caption{Rudin-Shapiro polynomials $P_m$ (solid curves) and $Q_m$ (dashed curves), for various orders $m$. (a), (b), (c), (d): $m=1, 2, 3, 4$, respectively (see Sec. \ref{Rud_Shap} for details).} 
\label{Figure2} 
\end{figure}

From (\ref{eq:Fbound}), it can be proved that the Fourier spectrum of the ARS sequence tends to a {\em constant} value in the infinite sequence limit \cite{Queffelec}, and is thus devoid of {\em any} strongly-localized spectral footprints. Again, this behavior resembles the ``flat-spectrum'' character of white-noise-like random sequences.
For the RS polynomials in (\ref{eq:PQ}) and (\ref{eq:PQ1}), a behavior similar to (\ref{eq:Fbound}) is found, with a {\em tighter} upper bound \cite{Cour}
\beq
\left|
P_m(\xi)
\right|\le 2^{\frac{m+1}{2}},~~
\left|
Q_m(\xi)
\right|\le 2^{\frac{m+1}{2}},
\label{eq:PQB}
\eeq
(compare with (\ref{eq:Fbound}), with $N=2^m$). The behavior of $P_m$ and $Q_m$ RS polynomials in a single period $0\le\xi<1$ is shown in Fig. \ref{Figure2}, for $1\le m\le4$. It is observed that, at any order, the two polynomial types exhibit {\em complementary} behaviors (maxima of $P_m$ correspond to minima of $Q_m$, and viceversa). Moreover, all the odd-order ($m=2l+1$) RS polynomials have their only zeros at $\xi=1/2$ ($P_m$-type) and $\xi=0, 1$ ($Q_m$-type) (see Figs. \ref{Figure2}(a) and \ref{Figure2}(c)), whereas even-order ($m=2l$) polynomials do not vanish for $0\le\xi<1$ (see Figs. \ref{Figure2}(b) and \ref{Figure2}(d)). Actually, it can be shown that, in general \cite{Cour},
\begin{subequations}
\begin{eqnarray}
P_{2l}(0)=P_{2l}(1/2)=2^l,~~P_{2l+1}(0)=2^{l+1},~~P_{2l+1}(1/2)=0,\\
Q_{2l}(0)=-Q_{2l}(1/2)=2^l,~~Q_{2l+1}(0)=0,~~Q_{2l+1}(1/2)=2^{l+1}.
\end{eqnarray}
\label{eq:PQSZ}
\end{subequations}

Concerning the BRS sequence in (\ref{eq:BRS}), its Fourier spectrum results from the superposition of a {\em flat} background and a {\em periodic} distribution ($\xi\in{\mathbb Z}$) of spectral peaks (see \cite{Queffelec}, \cite{Baake1} for details). This could be expected by noting that the BRS sequence in (\ref{eq:BRS}) can be interpreted as the superposition of an ARS sequence and a {\em periodic} sequence, whose cross-correlation can be proved to be zero \cite{Berthe}, \cite{Fogg}.
Quite remarkably, the BRS sequence exhibits {\em the same} Fourier spectrum as a Bernoulli {\em random} sequence in ${\cal A}_b=\left\{0,1\right\}$ whose symbols are chosen with equal probability of occurrence \cite{Baake1}, \cite{Baake2}; these two sequences are said to be {\em homometric}  \cite{Baake1} in technical jargon.

\section{Rudin-Shapiro Antenna Arrays}
\label{Array}
We now investigate in detail the radiation properties of representative classes of antenna arrays based on RS sequences. Referring to the geometry in Fig. \ref{Figure1}, the RS character is embedded in the array via the excitation amplitudes $I_n$ in (\ref{eq:AF}). Two cases are considered: a) ARS sequence ($I_n=\alpha_n$), and b) BRS sequence ($I_n=\beta_n$). 
Parametric studies have been performed, with emphasis on two meaningful array-oriented observables: the maximum directivity $D_M$ and the side-lobe-ratio (SLR) $\rho_S$, which can be compactly defined as
\beq
D_M=
\frac{2\left\|
F_p
\right\|^2_{\infty}}{\left\|
F_p
\right\|^2_2},
\label{eq:DM}
\eeq

\beq
\rho_S=\frac{\sup_{\theta \notin {\cal B}}|F_p(\theta)|^2}
{\left\|
F_p
\right\|^2_{\infty}}.
\label{eq:RSL}
\eeq
In (\ref{eq:DM}) and (\ref{eq:RSL}), $\left\|\cdot
\right\|_{\infty}$ and $\left\|\cdot\right\|_2$ denote the $L_{\infty}$ (supremum) and $L_2$ (rms) norms, respectively,
\beq
\left\|
F_p
\right\|^2_{\infty}=\sup_{-\pi/2\le \theta<\pi/2}|F_p(\theta)|^2,~~
\left\|
F_p
\right\|^2_2=
\int_{-\pi/2}^{\pi/2}|F_p(\theta)|^2\cos\theta d\theta,
\eeq
and ${\cal B}$ denotes the main lobe angular region. Using (\ref{eq:AF}), one can show that
\begin{subequations}
\beq
\left\|
F_p
\right\|^2_2=
2 c_0+4\sum_{\nu=1}^{N-1}c_{\nu}
\frac{\cos\left(k_0\nu d\eta\right)\sin\left(k_0\nu d\right)}{k_0\nu d},
\label{eq:Fp2}
\eeq
where the $c_{\nu}$ are correlation sums,
\beq
c_\nu=
\displaystyle{\sum_{l=0}^{N-1-\nu}}I_l I_{l+\nu}.
\label{eq:cnu}
\eeq
\label{eq:Fc}
\end{subequations}

This provides the background for the theoretical considerations and synthetic numerical experiments that follow.

\subsection{Flat-Spectrum ``Omnidirectional'' Arrays}
\label{Flat}
\begin{figure} 
\begin{center}
\includegraphics[width=10cm]{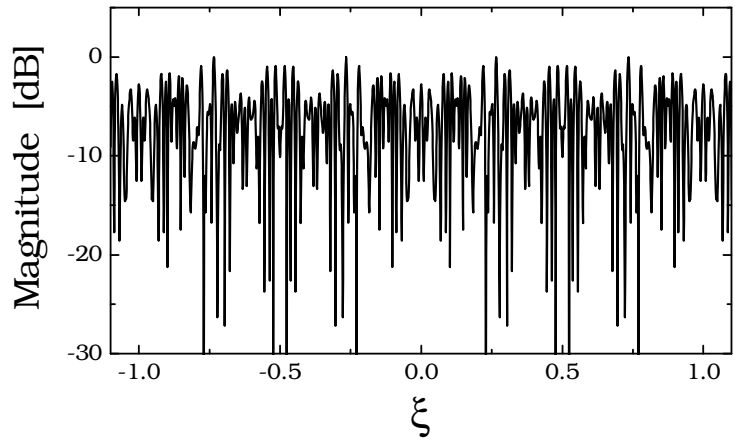}
\end{center}
\caption{ARS antenna array with $N=100$ elements. Radiation spectrum $|F_s(\xi)|^2$ (normalized to its maximum value) vs. spectral parameter $\xi$ (see (\ref{eq:AF})). Maximum directivity is $D_M\approx5.7$dB, almost independent of the inter-element spacing and phasing.} 
\label{Figure3} 
\end{figure}

\begin{figure} 
\begin{center}
\includegraphics[width=14cm]{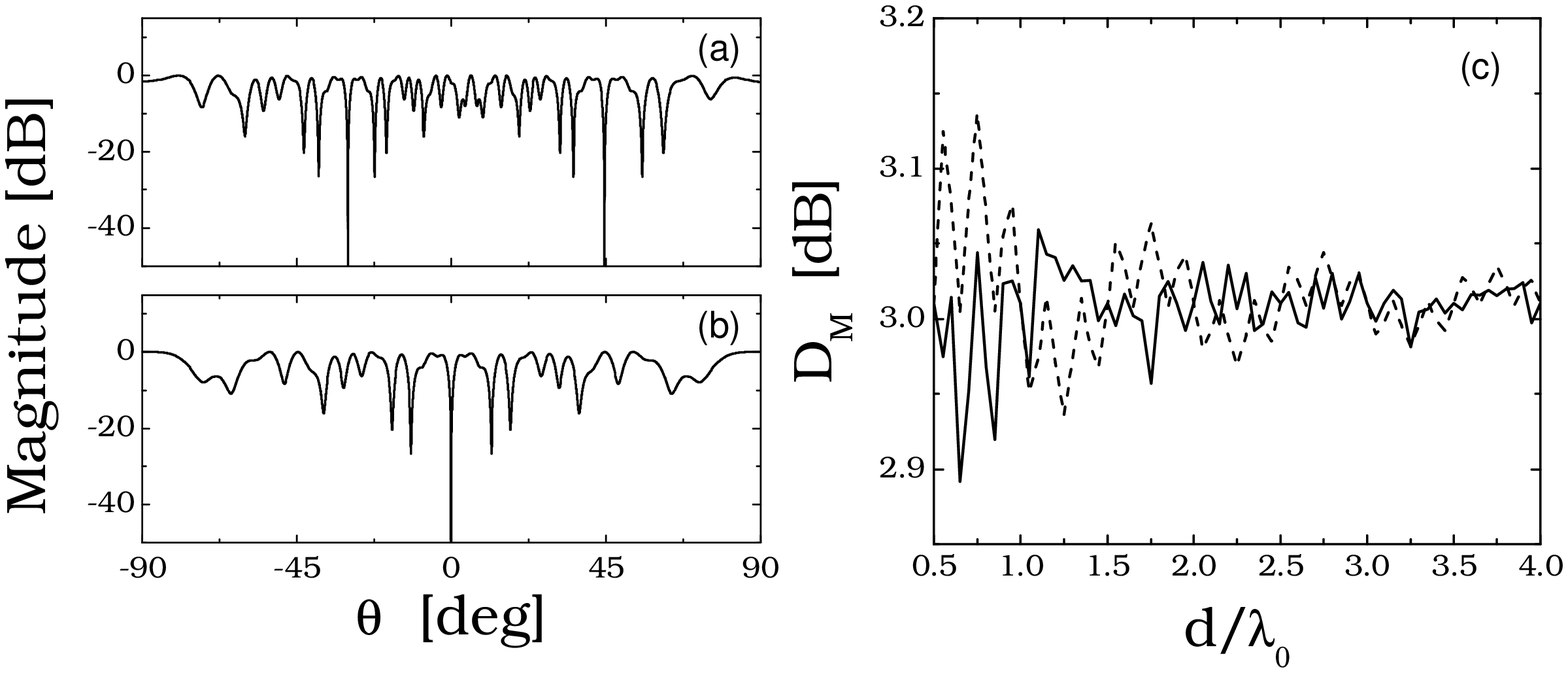}
\end{center}
\caption{Examples of antenna arrays synthesized using the RS polynomials $P_5$ and $Q_5$ ($N=32$ elements). ~~(a):  Radiation pattern $|F_p(\theta)|^2$ (normalized to its maximum value) for $P_5$-type array, with $d=0.83\lambda_0$ and $\eta=0.1$ (with prescribed nulls at $\theta=-30^o$ and $\theta=45^o$), having maximum directivity $D_M=2.9$dB. ~~(b): Radiation pattern for $Q_5$-type array, with $d=0.5\lambda_0$ and $\eta=0$ (with prescribed null at $\theta=0$), having maximum directivity $D_M=3$dB. 
~~(c): Maximum directivity $D_M$ vs. inter-element spacing for the two configurations in (a) and (b), shown as solid and dashed curves, respectively.} 
\label{Figure4} 
\end{figure}

We begin with the class of antenna arrays based on the {\em ARS sequence} $\left\{\alpha_n\right\}$ in (\ref{eq:ARS}).
A typical example of the radiation spectrum is shown in Fig. \ref{Figure3} for a 100-element array. As could be expected from the spectral properties of the ARS sequence discussed in Sec. \ref{Rud_Shap}, the radiation spectrum exhibits a fairly {\em flat} behavior (no distinct localized spectral peaks) with truncation-induced rapid oscillations. This results in a radiation pattern with globally ``omnidirectional'' characteristics. Interestingly, for this class of arrays, one can establish some general theoretical bounds on the maximum directivity $D_M$, which are practically independent of the inter-element spacing and phasing. First, one can show that the radiation pattern satisfies the constraint
\beq
\left\|
F_p
\right\|^2_{\infty}\le (2+\sqrt{2})^2 N,~~
\left\|
F_p
\right\|^2_2
\approx 2c_0=2N.
\label{eq:ARS_AR}
\eeq
In (\ref{eq:ARS_AR}), the upper bound on the $L_{\infty}$ norm follows directly from (\ref{eq:Fbound}). The approximation for the $L_2$ norm, numerically verified {\em a posteriori} (see below), is obtained from (\ref{eq:Fp2}) by neglecting the $\nu$-summation; this term vanishes exactly for $d=p\lambda_0/2, p\in {\mathbb N}$, and is more generally negligible since -- in view of (\ref{eq:corr}) and (\ref{eq:cnu}), for arrays of {\em finite} but moderate-to-large size -- one expects $|c_{\nu}|\ll c_0$, for $\nu\ge1$. Accordingly, it follows from (\ref{eq:DM}) that the bound on the maximum directivity $D_M$ is given by
\beq
D_M\lesssim (2+\sqrt{2})^2 \approx 10.7~\mbox{dB}.
\label{eq:DMB}
\eeq
By comparison with the computed value (via (\ref{eq:DM})) in Fig. \ref{Figure3} (which is $D_M\approx5.7$, practically independent of $d/\lambda_0$ and $\eta$), this {\em general} bound turns out to be rather {\em loose}. In this connection, much tighter bounds can be established for arrays based on the RS polynomials in (\ref{eq:PQ}). In view of (\ref{eq:PQ1}), these arrays can be synthesized from the array factor $F_s(\xi)$ in (\ref{eq:AF}) with $I_n=p_n, q_n$ and $N=2^m$, yielding radiation patterns of the type
\beq
|F_p(\theta)|^2=\left\{
\begin{array}{ll}
|P_m[d(\sin\theta-\eta)/\lambda_0]|^2,\\
|Q_m[d(\sin\theta-\eta)/\lambda_0]|^2.
\end{array}
\right.
\eeq
For this class of arrays, the relationships in (\ref{eq:ARS_AR}) can be rewritten as
\beq
\left\|
F_p
\right\|^2_{\infty}\le 2^{m+1},~~
\left\|
F_p
\right\|^2_2\approx 2\cdot2^m,
\label{eq:ARS_AR1}
\eeq
where the upper bound on the $L_{\infty}$ norm now follows from (\ref{eq:PQB}), whereas the approximation for the $L_2$ norm is obtained and justified as before (see (\ref{eq:ARS_AR}), with $N=2^m$). This yields the following bound on the maximum directivity
\beq
D_M\lesssim 2\approx 3~\mbox{dB},
\label{eq:DMB1}
\eeq
which is consequently {\em tighter} than the general bound in (\ref{eq:DMB}). By this synthesis, one could therefore obtain an antenna array with nearly ``omnidirectional'' characteristics. 
Moreover, for the special case of odd-order ($m=2l+1$) RS polynomials, by exploiting the two available degrees of freedom ($d/\lambda_0$ and $\eta$) to scale and shift the spectral zeros in (\ref{eq:PQSZ}), one can place up to two nulls at arbitrary directions, while keeping the {\em global} omnidirectional character. Two examples are illustrated in Fig. \ref{Figure4}, in connection with 5-th-order RS polynomials ($N=32$ antenna elements). Specifically, Fig. \ref{Figure4}(a) shows the radiation pattern synthesized with a $P_5$ polynomial, with array parameters ($d/\lambda_0=0.83$ and $\eta=0.1$) chosen so as to place two nulls at $\theta=-30^o$ and $\theta=45^o$. Figure \ref{Figure4}(b) shows the radiation pattern synthesized with a $Q_5$ polynomial, with array parameters ($d/\lambda_0=0.5$ and $\eta=0$) chosen so as to place a null at broadside ($\theta=0$).
The maximum directivity vs. the inter-element spacing is shown in Fig. \ref{Figure4}(c) for both configurations. An essentially {\em flat} behavior (magnified by the expanded scale chosen in the plot) is observed, with computed values very close to the theoretical bound in (\ref{eq:DMB1}); this also serves as an indirect {\em a posteriori} validation for the approximation of the $L_2$ norm used in (\ref{eq:ARS_AR}) and (\ref{eq:ARS_AR1}). 
Clearly, in view of the omnidirectional character, with the absence of a main radiation beam, the SLR parameter $\rho_S$ in (\ref{eq:RSL}) is not of interest for this class of arrays.

\subsection{Thinned Arrays}
\label{Thinned}
\begin{figure} 
\begin{center}
\includegraphics[width=10cm]{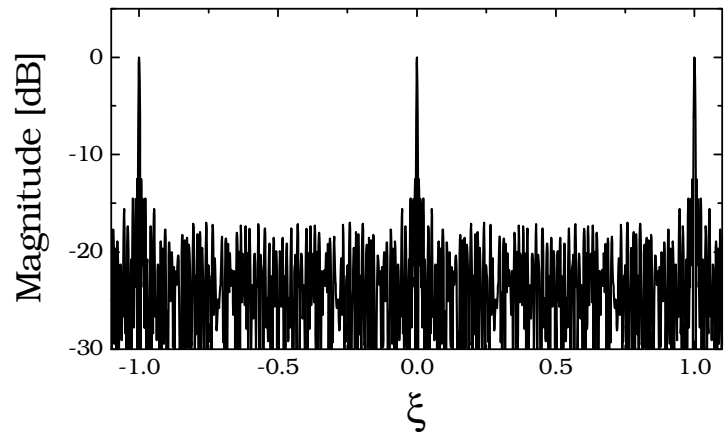}
\end{center}
\caption{Thinned BRS antenna array with $N=200$ ($N_a=90$ active elements). Radiation spectrum $|F_s(\xi)|^2$ (normalized to its maximum value) vs. spectral parameter $\xi$.} 
\label{Figure5} 
\end{figure}

Next, we move on to antenna arrays based on the {\em BRS sequence} $\left\{\beta_n\right\}$ in (\ref{eq:BRS}).
Assuming $I_n=\beta_n$ in the array factor in (\ref{eq:AF}), and recalling that $\beta_n\in{\cal A}_b=\left\{0,1\right\}$ (i.e., antenna elements are either ``off'' or ``on''), one may interpret this choice as a special {\em deterministic} ``array-thinning'' strategy \cite{Mailloux}. Recalling the spectral properties of BRS sequences summarized at the end of Sec. \ref{Rud_Shap}, one can expect a radiation spectrum constituted of an ARS-type {\em flat} background (see Fig. \ref{Figure3}) with superposed {\em discrete} spectral peaks (grating lobes) typical of {\em periodic} arrays. These features are clearly visible in the example shown in Fig. \ref{Figure5} for an array with $N=200$ featuring $N_a=90$ ``active'' (i.e., $I_n=1$) elements.
Based on these premises, and recalling the symbol equi-occurrence property for infinite RS sequences, one may anticipate radiation properties (in terms of beamwidth, directivity and SLR) comparable to those of a fully-populated {\em periodic array}, using {\em nearly half} of the elements, at the expense of a slight pattern degradation (due to the flat background spectrum) which is however expected (and acceptable) in most typical array-thinning strategies \cite{Mailloux}. In other words, for a {\em fixed} number of active antenna elements $N_a$ and array aperture $L_A$ (defined as the distance between the first and last active elements), BRS-type arrays could yield an increase by a factor $\sim 2$ in the maximum value allowable for the
average inter-element spacing $d_{av}=L_A/(N_a-1)$ so as to avoid the appearance, in the visible range, of grating lobes ($d_{av}\sim 2\lambda_0$ vs. $d_{av}=\lambda_0$ for periodic arrays). 

\begin{figure} 
\begin{center}
\includegraphics[width=14cm]{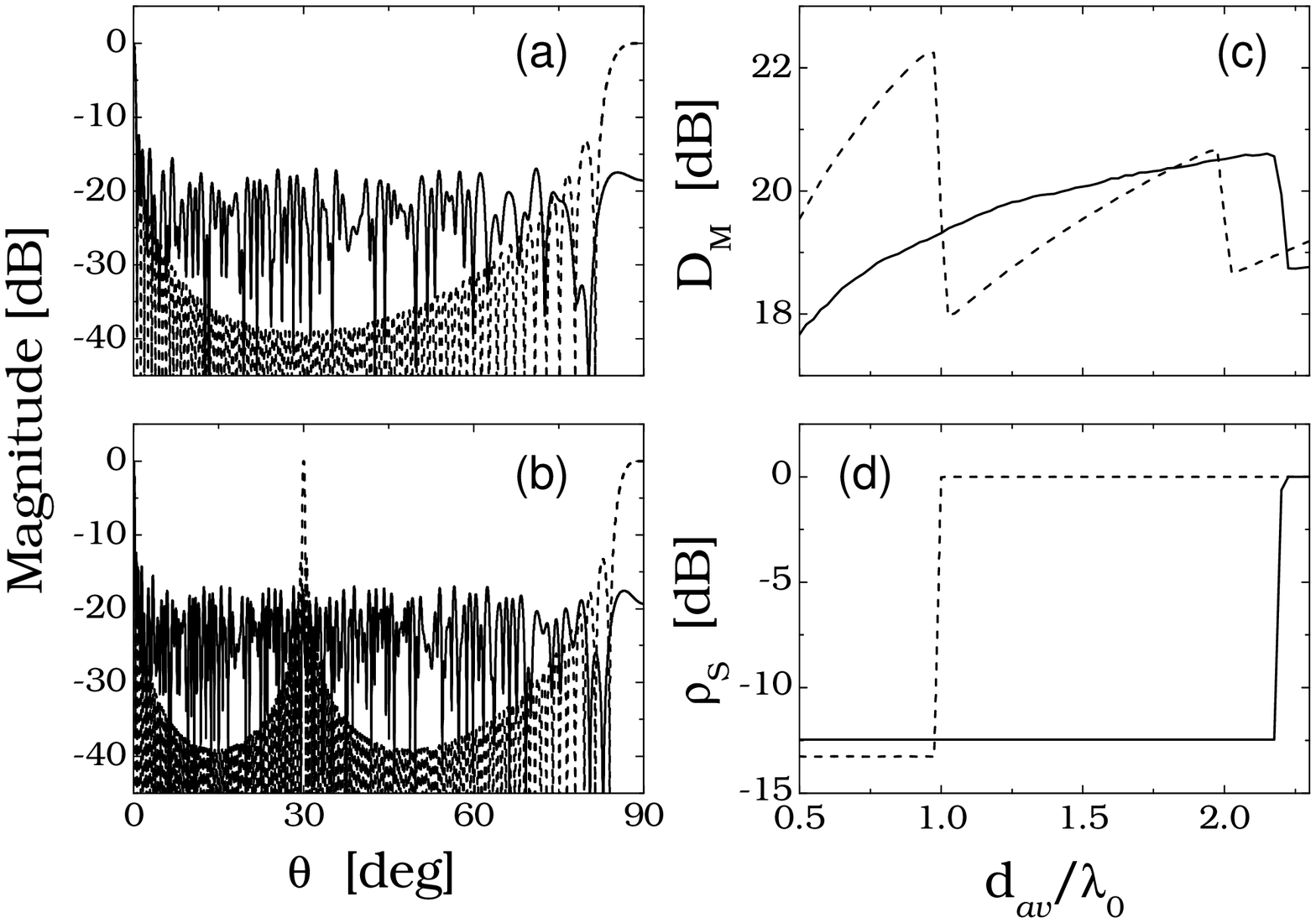}
\end{center}
\caption{As in Fig. \ref{Figure5}, but radiation characteristics for $\eta=0$ (solid curves). (a), (b): Radiation patterns $|F_p(\theta)|^2$ (normalized to their maximum value) for average inter-element spacings $d_{av}=\lambda_0$ and $d_{av}=2\lambda_0$, respectively. (c), (d): Maximum directivity $D_M$ in (\ref{eq:DM}) and side-lobe ratio $\rho_S$ in (\ref{eq:RSL}), respectively, vs. average inter-element spacing. Also shown (dashed), for comparison, is the reference behavior of a periodic array of the same size.} 
\label{Figure6} 
\end{figure}

Figure \ref{Figure6} illustrates the radiation characteristics of a typical BRS-type array with $N_a=90$ active elements. Specifically, Figs. \ref{Figure6}(a) and \ref{Figure6}(b) show the radiation patterns for average inter-element spacings $d_{av}=\lambda_0$ and $d_{av}=2\lambda_0$, respectively. Also shown, for comparison, is the reference behavior of a periodic array with same number of elements and inter-element spacing (and, hence, comparable beamwidth). As expected, while the periodic array displays the well-known grating lobes, the BRS-array {\em does not}. The expected slight degradation in the radiation pattern (due to the flat background spectrum) is observed with amplitude below the secondary lobe level ($\sim -13$dB). This is better quantified in Figs. \ref{Figure6}(c) and \ref{Figure6}(d), where the maximum directivity $D_M$ and the SLR $\rho_S$ vs. the average inter-element spacing are shown. It is observed that the BRS-array characteristics are only slightly worse than those of the periodic array for $d_{av}<\lambda_0$, and become almost uniformly better beyond the critical value $d_{av}=\lambda_0$ at which the periodic array characteristics experience an abrupt degradation due to the entrance in the visible range of the first grating lobe (see Fig. \ref{Figure6}(a)). This trend holds up to values of $d_{av}\approx 2.2 \lambda_0$, at which the first grating lobe of the BRS array enters the visible range. Qualitatively similar behaviors have been observed by us for arrays of different size. Some representative results are summarized in Table I in terms of maximum directivity and SLR for the case $d_{av}=\lambda_0$ and $\eta=0$. It is observed that, with the exception of small-size arrays ($N_a\lesssim 50$), the SLR values are always comparable with the reference periodic-array level ($\sim -13$dB).

\section{Conclusions and Perspectives}
\label{Conclusions}
This paper has dealt with the radiation properties of 1-D quasirandom antenna arrays based on Rudin-Shapiro (RS) sequences. Potential applications have been suggested, based on known spectral and statistical properties  of these sequences. Results so far seem to indicate that RS sequences might provide new perspectives for the synthesis of antenna arrays.

In particular, the ARS-type synthesis described in Sec. \ref{Flat} might be of potential interest for smart reconfigurable antenna systems where {\em omnidirectional} receiving/transmitting features are needed, as a possible alternative to the code-based synthesis approaches in \cite{Code}. Other potentially interesting applications could be envisaged in the area of reflect-array synthesis of ``simulated corrugated surfaces'' \cite{Stephen} and, more generally, ``virtual shaping'' applications in radar countermeasures \cite{Swandic}. In this connection, use could be made of recent advances in the practical synthesis of artificial impedance surfaces \cite{IEEE_SI} for designing 1-D (strip) or 2-D (patch) planar arrangements based on suitable ``dual'' impedance surfaces. Examples are perfect-electric and perfect-magnetic conductors capable of providing the opposite-sign reflection coefficients which, in a first-order physical optics approximation of the relevant scattering problem, play the role of the ARS excitation amplitudes in the array factor here. Such structures, currently under investigation, could yield strong reduction of specular reflection and {\em uniform} angular distribution of the backscattered power.

On the other hand, the BRS-type synthesis described in Sec. \ref{Thinned} seems to offer a simple {\em fully deterministic} alternative to typical {\em random}
thinning strategies \cite{Mailloux} for arrays of moderate-to-large size.
In this connection, it is interesting to observe that, in view of the {\em homometry} properties mentioned at the end of Sec. \ref{Array}, this strategy is in a sense {\em equivalent} to a {\em completely random} scheme, in which the antenna elements are turned on/off with equal probability.

Planned future investigations involve the extension to 2-D geometries, as well as applications to reflect-array synthesis for radar cross-section reduction and control.
Exploration of the radiation properties of array configurations based on other categories of aperiodic sequences continues.

\section*{Acknowledgements}
L.B. Felsen acknowledges partial support from Polytechnic University, Brooklyn, NY 11201 USA.


\newpage
\begin{table}[h]
\begin{center}
\caption{BRS antenna array as in Fig. \ref{Figure6}. Maximum directivity $D_M$ and SLR $\rho_S$ for $d_{av}=\lambda_0, \eta=0$, and various values of the number of active elements $N_a$. The SLR values obtained from numerical simulation were almost independent of $d_{av}$ and $\eta$.}
\bigskip
  \begin{tabular}{|c|c|c|}
  \hline \hline
    $N_a$ & $D_M$ [dB] &$\rho_s$ [dB]\\  
\hline \hline
10 & 9.67 & -5.9\\
\hline
25 & 13.6 & -8.8\\
\hline
50 & 16.8 & -11.9\\
\hline
100 & 19.8 &-11.2\\
\hline
250 & 23.8 &-13.8\\
\hline
500 & 26.9 & -12.3\\
\hline
\end{tabular}
\end{center}
\label{TableI}
\end{table}

\newpage
%

\newpage
%

\newpage
%

\newpage
%

\newpage
%

\newpage
%


\begin{thebibliography}{999}

\bibitem{Tiling_AP}{V. Pierro, V. Galdi, G. Castaldi, I. M. Pinto, and L. B. Felsen, ``Radiation properties of planar antenna arrays based on certain categories of aperiodic tilings,'' {\em IEEE Trans. Antennas Propagat.}, vol. 53, No. 2, pp. 635--644, Feb. 2005.}

\bibitem{Fibonacci_AP}{V. Galdi, G. Castaldi, V. Pierro, I. M. Pinto, and L. B. Felsen, ``Parameterizing quasiperiodicity: Generalized Poisson summation and its application to modified-Fibonacci antenna arrays,'' to be published in {\em IEEE Trans. Antennas Propagat.}, vol. 53, No. 6, June 2005.}

\bibitem{Shechtman}{D. Shechtman, I. Blech, D. Gratias, and J. W. Cahn, ``Metallic phase with long-range orientational order and no translation symmetry,'' {\em Phys. Rev. Letts.}, vol. 53, No. 20, pp. 1951--1953, Nov. 1984.}

\bibitem{Levine}{D. Levine and P. J. Steinhardt, ``Quasicrystals: A new class of ordered structures,'' {\em Phys. Rev. Letts.}, vol. 53, No. 26, pp. 2477--2480, Dec. 1984.}

\bibitem{Grunbaum}{B. Gr\"unbaum and G. C. Shepard, {\em Tilings and Patterns}. New York (NY): Freeman, 1987.}

\bibitem{Senechal1}{M. Senechal, {\em Quasicrystals and Geometry}. Cambridge (UK): Cambridge University Press, 1995.}

\bibitem{Shapiro1}{H. S. Shapiro, {\em Extremal Problems for Polynomials and Power Series}. Master thesis,
Massachusetts Institute of Technology, USA, May 1951.}

\bibitem{Rudin1}{W. Rudin, ``Some theorems on Fourier coefficients,'' {\em Proc. Amer. Math. Soc.}, vol. 10, pp. 855–-859, 1959.}

\bibitem{Brillhart1}{J. Brillhart and P.Morton, ``A case study in mathematical research: The
Golay-Rudin-Shapiro sequence,'' {\em Amer. Math. Mon.}, vol. 103, No. 10, pp. 854--869, Dec. 1996.}

\bibitem{Dixon}{R. C. Dixon, {\em Spread Spectrum Systems with Commercial Applications}. New York (NY): Wiley, 1994.}

\bibitem{Cour}{A. la Cour-Harbo, ``On the Rudin-Shapiro transform,'' accepted for publication in {\em Appl. Comp. Harmonic Analysis} (preprint available at http://www.control.auc.dk/$\sim$alc/Publications/acha-art-preprint.pdf).}

\bibitem{Vasco}{M. S. Vasconcelos and E. L. Albuquerque, ``Transmission fingerprints in quasiperiodic multilayers,'' {\em Phys. Rev. B}, vol. 59, No. 17, pp. 11128-–11131, May 1999.}

\bibitem{Mailloux}{R. J. Mailloux, {\em Phased Array Antenna Handbook}. Boston (MA): Artech House, 1994.}

\bibitem{Queffelec}{M. Queff\'elec, {\em Substitution Dynamical Systems - Spectral Analysis}, Lecture Notes in Mathematics, vol. 1294. Berlin: Springer, 1987.}

\bibitem{Luck}{C. Godr\`eche and J. M. Luck, ``Multifractal analysis in reciprocal space and the nature of the Fourier transform of self-similar structures,'' {\em J. Phys. A: Math. Gen.}, vol. 23, No 16, pp. 3769--3797, Aug. 1990.}  

\bibitem{Axel}{F. Axel, J.-P. Allouche, and Z.-Y. Wen, ``On certain properties of high-resolutions x-ray diffraction spectra of finite-size generalized Rudin-Shapiro multilayer heterostructures,'' {\em J. Phys.: Condens. Matter}, vol. 4, No. 45, pp. 8713--8728, Nov. 1992.}

\bibitem{Berthe}{V. Berth\'e, ``Conditional entropy of some automatic sequences,''
{\em J. Phys. A: Math. Gen.}, vol. 27, No. 24, pp. 7993--8006, Dec. 1994.}

\bibitem{Fogg}{N. P. Fogg, V. Berth\'e, S. Ferenczi, C. Mauduit, A. Siegel (Eds.),
{\em Substitutions in Dynamics, Arithmetics, and Combinatorics}, Lecture Notes in Mathematics, vol. 1794. Berlin: Springer, 2002.}

\bibitem{Brillhart}{J. Brillhart, ``On the Rudin-Shapiro polynomials,'' {\em Duke Math. J.}, vol. 40, No. 2, pp. 335-–353, June 1973.}

\bibitem{Golay}{M. J. E. Golay, ``Complementary series,'' {\em IEEE Trans. Information Theory}, vol. 7, No. 2, pp. 82-–87, Apr. 1961.}

\bibitem{Allouche}{J. P. Allouche and P. Liardet, ``Generalized Rudin-Shapiro sequences,'' {\em Acta Arithmetica}, vol. 60, pp. 1--27, 1991/1992.}

\bibitem{Benke}{G. Benke, ``Generalized Rudin-Shapiro systems,'' {\em J. Fourier Anal. Appl.}, vol. 1, No. 1, pp. 87–-101, Jan. 1994.}

\bibitem{Baake1}{M. H\"offe and M. Baake, ``Surprises in diffuse scattering,'' {\em Zeitschrift f\"ur Kristallographie}, vol. 215, No. 8, pp. 441--444, Aug. 2000.} 

\bibitem{Baake2}{M. H\"offe and M. Baake, ``Diffraction of random tilings: Some rigorous results,'' {\em J. Stat. Phys.}, vol. 99, No. 1/2, pp. 219--261, Apr. 2000.} 

\bibitem{Code}{S. E. El-Khamy, O. A. Abdel-Alim, A. M. Rushdi, and A. H. Banah, ``Code-fed omnidirectional arrays,''
{\em IEEE J. Oceanic Eng.}, vol. 14, No. 4, pp. 384--395, Oct. 1989.}

\bibitem{Stephen}{D. S. Stephen, T. Mathew, K. A. Jose, C. K. Aanandan, P. Mohanan, and K. G. Nair, ``New Simulated corrugated scattering surface giving wideband characteristics,'' {\em Electronic Letts.}, vol. 29, No. 4, pp. 329--331, Feb. 1993.}

\bibitem{Swandic}{J. R. Swandic, ``Bandwidth limits and other considerations for monostatic RCS reduction by virtual shaping,'' Tech. Report. No. A927224, Naval Surface Warfare Center, Carderock Div., Bethesda MD, Jan. 2004.} 

\bibitem{IEEE_SI}{{\em IEEE Trans. Antennas and Propagation}, Special Issue on ``Artificial Magnetic Conductors, Soft/Hard Surfaces, and other Complex Surfaces,'' P.-S. Kildal, A. Kishk, and S. Maci, Guest Eds., 2005 (in print).}


\end{thebibliography}
\end{document}